\documentclass[11pt]{article}
\pdfoutput=1 
\usepackage{amssymb}
\usepackage{amsmath}
\usepackage{amstext}
\usepackage{graphicx}
\usepackage{epsfig}
\usepackage{verbatim} 
\usepackage{fancybox}
\usepackage{color}
\usepackage{mathdots}
\usepackage{anyfontsize}
\usepackage{ulem}
\usepackage{enumitem}
\usepackage{caption}
\usepackage{amsthm}
\usepackage{subfig}
\usepackage{mathtools}
\usepackage{youngtab}
\usepackage{bbm}
\usepackage{parskip}
\usepackage[numbers,sort&compress]{natbib}
\usepackage{ytableau}
\usepackage[utf8]{inputenc}
\usepackage{tikz}
\usepackage{simplewick}
\usetikzlibrary{positioning, trees,decorations, decorations.markings, decorations.pathmorphing, arrows, shapes.geometric}

\newcommand{\Comment}[1]{{}}
\definecolor{darkblue}{rgb}{0.15,0.35,0.55}
\definecolor{reddish}{rgb}{0.65, 0.2, 0.2}
\usepackage[linktocpage=true]{hyperref}
\hypersetup{
colorlinks=true,
citecolor=darkblue,
linkcolor=reddish,
urlcolor=darkblue,
pdfauthor={},
pdftitle={},
pdfsubject={}
}

\usepackage{cleveref}

\newcommand{\be}{\begin{equation}}
\newcommand{\ee}{\end{equation}}
\newcommand{\nn}{\nonumber}
\newcommand{\intM}{\int_{\mathcal{M}} }
\newcommand{\norm}[1]{\left\lVert#1\right\rVert}

\numberwithin{equation}{section}

\makeatletter
\def\thickhline{%
  \noalign{\ifnum0=`}\fi\hrule \@height \thickarrayrulewidth \futurelet
   \reserved@a\@xthickhline}
\def\@xthickhline{\ifx\reserved@a\thickhline
               \vskip\doublerulesep
               \vskip-\thickarrayrulewidth
             \fi
      \ifnum0=`{\fi}}
\makeatother

\newlength{\thickarrayrulewidth}
\begin{document}

\renewcommand{\thefootnote}{\fnsymbol{footnote}}
~
\vspace{1.75truecm}
\thispagestyle{empty}
\begin{center}
{\LARGE \bf{
Bootstrapping Closed Hyperbolic Surfaces 
}}
\end{center} 

\vspace{1cm}
\centerline{\Large James Bonifacio\footnote{\href{mailto:james.j.bonifacio@gmail.com}{\texttt{james.j.bonifacio@gmail.com}}}
}
\vspace{.5cm}

\centerline{{\it Department of Applied Mathematics and Theoretical Physics,}}
 \centerline{{\it   University of Cambridge, Wilberforce Road, Cambridge CB3 0WA, U.K.}} 
 \vspace{.25cm}
 
\vspace{1cm}
\begin{abstract}
\noindent
The eigenvalues of the Laplace--Beltrami operator and the integrals of products of eigenfunctions and holomorphic $s$-differentials satisfy certain consistency conditions on closed hyperbolic surfaces. These consistency conditions can be derived by using spectral decompositions to write quadruple overlap integrals in terms of triple overlap integrals in different ways. We show how to efficiently construct these consistency conditions and use them to derive upper bounds on eigenvalues, following the approach of the conformal bootstrap. As an example of such a bootstrap bound, we find a numerical upper bound on the spectral gap of closed orientable hyperbolic surfaces that is nearly saturated by the Bolza surface. 
\end{abstract}

\newpage

\setcounter{tocdepth}{2}
\tableofcontents
\renewcommand*{\thefootnote}{\arabic{footnote}}
\setcounter{footnote}{0}

\section{Introduction}
Hyperbolic surfaces are important objects in several areas of physics and mathematics, including quantum gravity, number theory, and topology.
 We can write the Riemann tensor on a hyperbolic surface $(\mathcal{M}, \hat{g})$ as
\be
R_{pqrs} = \kappa \left( \hat{g}_{pr} \hat{g}_{qs} - \hat{g}_{ps} \hat{g}_{qr} \right),
\ee
where the negative constant $\kappa$ is the Gaussian curvature, which we henceforth normalise to $-1$.
A well-established approach to investigating closed hyperbolic surfaces is through the spectral theory of the Laplace--Beltrami operator $\Delta$---see, for example, Ref.~\cite{buser1992} for an introduction to this subject. The Laplace--Beltrami operator on a closed hyperbolic surface is a nonnegative elliptic operator with eigenfunctions $\phi_i$,  $i=0, 1, 2, \dots$, which satisfy the eigenvalue equation $\Delta \phi_i = \lambda_i \phi_i$. The discrete eigenvalues $\lambda_i$ satisfy
\be
0=\lambda_0 < \lambda_1 \leq \lambda_2 \leq \dots \rightarrow \infty,
\ee
where we assume that the surface is connected so that $\lambda_0$ is the unique zero eigenvalue.

An important spectral quantity is the first nonzero eigenvalue $\lambda_1$, which is called the spectral gap. Various bounds are known on the spectral gap. For example, the Yang--Yau bound  \cite{YangYau1980}, with an improvement pointed out in Ref.~\cite{ElSoufi83}, implies that on a closed orientable hyperbolic surface of genus $g$
\be \label{eq:YangYau}
\lambda_1 \leq \frac{ 2 \lfloor (g+3)/2 \rfloor}{g-1}.
\ee
In particular, this implies that $\lambda_1  \leq 4$, since $g \geq 2$ for a closed hyperbolic surface. It has been conjectured that the maximum value of the spectral gap is achieved by the Bolza surface \cite{Strohmaier_2012}, which is a genus-2 surface with $\lambda_1 \approx 3.839$. This surface is the closed genus-2 surface with the largest symmetry group and can be obtained by identifying opposite sides of a regular octagon in the hyperbolic disk. The Bolza surface has been studied by physicists in the context of quantum chaos \cite{Aurich1989} (see also Ref.~\cite{Afkhami-Jeddi:2021qkf} for recent results on chaos in the spectra of genus-3 Riemann surfaces). Recently, a bound stronger than the Yang--Yau bound was obtained for closed hyperbolic surfaces with $g=3$ \cite{ros2021}, while improved bounds were found for almost all other genera in Ref.~\cite{karpukhin2021}. It has long been known that the spectral gap is bounded above by a constant that approaches $1/4$ as $g \rightarrow \infty$ \cite{Huber1974}, but it was only recently shown that there exists a sequence of surfaces with $\lambda_1 \rightarrow 1/4$ and $g\rightarrow \infty$ \cite{hide2021near}. While $\lambda_1$---and, more strongly, $\lambda_{2g-3}$---can be arbitrarily small \cite{Buser1977}, there does exist a nonzero lower bound on the \textit{generic} value of $\lambda_1$ as $g \rightarrow \infty$. For example, for any $\epsilon>0$, the probability that a surface has $\lambda_1 > 3/16- \epsilon$ goes to one as $g \rightarrow \infty$, according to the normalised Weil--Petersson measure on moduli space \cite{lipnowski2021, wu2021}.

One way to obtain bounds on eigenvalues and integrals of products of eigenfunctions is using a bootstrap approach \cite{Bonifacio:2020xoc, Bonifacio:2021msa}. This approach is inspired by the conformal bootstrap for conformal field theories (CFTs) \cite{Rattazzi:2008pe, Rychkov:2009ij} (see Ref.~\cite{Poland:2018epd} for a review of the numerical conformal bootstrap).\footnote{A different mathematical context in which bootstrap ideas are useful is sphere packing \cite{Hartman:2019pcd, Afkhami-Jeddi:2020hde}.} These bounds have an interpretation as constraints on the masses and coupling constants of Kaluza--Klein (KK) modes in theories with compact extra dimensions. As explained below, these bounds are obtained from certain consistency conditions that can be derived by using spectral decompositions to write quadruple overlap integrals of derivatives of Laplacian eigenmodes in different ways.\footnote{Spectral decompositions have also recently been shown to be useful in the context of the modular bootstrap \cite{Benjamin:2021ygh}.} For Ricci-flat manifolds, these consistency conditions can also be obtained by considering amplitudes of KK gravitons \cite{Bonifacio:2019ioc}.\footnote{See also Ref.~\cite{DeLuca:2021ojx} for some recent bounds on Laplacian eigenvalues with motivation from Kaluza--Klein theory.} In Ref.~\cite{Bonifacio:2021msa}, consistency conditions for closed hyperbolic manifolds in general dimensions were obtained from integrals of products of a scalar eigenfunction with up to 16 derivatives. In this paper, we show how to go much further by specialising to two dimensions. By using helicity tensors, we can derive consistency conditions for closed hyperbolic surfaces coming from quadruple overlap integrals with arbitrarily many derivatives and involving both eigenfunctions and holomorphic $s$-differentials. These are analogous to the crossing equations for mixed correlators and correlators of spinning bosonic operators in the conformal bootstrap \cite{Kos:2014bka, Dymarsky:2017xzb, Dymarsky:2017yzx}.

Once we have a set of consistency conditions, we can use them to derive bounds on eigenvalues and integrals of products of eigenfunctions by following the approach of the conformal bootstrap. For example, in Ref.~\cite{Bonifacio:2021msa} we obtained the following bound for consecutive nonzero eigenvalues of the Laplace--Beltrami operator on closed orientable hyperbolic surfaces:
 \be
 \lambda_{i+1} \leq 1/2+3 \lambda_i+\sqrt{\lambda_i^2+2 \lambda_i+1/4}, \quad i \geq 1.
\ee
When we consider consistency conditions coming from integrals of products of holomorphic $s$-differentials, it also becomes possible to find an upper bound on the spectral gap. For example, in this work, we show that by considering consistency conditions coming from the quadruple overlap integrals of two distinct holomorphic differentials we can obtain the numerical upper bound
\be \label{eq:genus-2-bound}
\lambda_1 \leq 3.8388977,
\ee
which is remarkably close to the spectral gap of the Bolza surface $\lambda_1 \approx 3.8388873$ \cite{Strohmaier_2012}.
This provides evidence for the conjecture that the Bolza surface has the largest spectral gap of any closed hyperbolic surface \cite{Strohmaier_2012}. As this work was nearing completion, we became aware of an upcoming paper that has some overlap with ours \cite{Kravchuk:2021akc}, including a version of the bound in Eq.~\eqref{eq:genus-2-bound} that is verified using exact rational arithmetic. We recommend reading Ref.~\cite{Kravchuk:2021akc} as well. 

The outline of the rest of this paper is as follows: in Section \ref{sec:2Dtensors}, we review a formalism for describing tensors of definite helicity in two dimensions and discuss tensor spectral decompositions and triple overlap integrals. In section \ref{sec:ccs}, we show how this formalism makes it easier to derive consistency conditions coming from quadruple overlap integrals involving multiple eigenfunctions and holomorphic $s$-differentials. In Section \ref{sec:bounds}, we give some examples of bootstrap bounds on the spectral gap that can be derived using these consistency conditions. We conclude in Section \ref{sec:Conclusions}.
 
\section{Tensors in two dimensions}
\label{sec:2Dtensors}
We begin by reviewing some properties of helicity tensors on two-dimensional closed hyperbolic manifolds. See Refs.~\cite{Alvarez:1982zi, dHokerPhong1986, dHokerPhong88} for uses of this formalism in the context of string perturbation theory. 

\subsection{Topology and coordinates}
We consider a closed hyperbolic manifold $(\mathcal{M}, \hat{g})$ with curvature $\kappa=-1$, where $\hat{g}_{pq}$ is the metric. Our manifolds are also always assumed to be smooth, orientable, and connected.
Topologically, a closed surface is classified by its genus $g\in \mathbb{Z}_{\geq 0}$. Closed surfaces with $g \geq 2$ admit hyperbolic metrics and on such surfaces there is a moduli space of hyperbolic structures, i.e., hyperbolic metrics modulo diffeomorphisms, of real dimension $6(g-1)$.  By the Gauss--Bonnet theorem, the volume of a hyperbolic surface of genus $g$ is $V= 4 \pi (g-1)$.

In a neighbourhood of every point, there exist isothermal coordinates $x_p$, $p=1,2$, such that the line element takes the form
\be
ds^2 = e^{2 \sigma} \delta_{pq} dx^p dx^q. 
\ee
The Ricci scalar in these coordinates is $R = -2 e^{-2\sigma} (\partial_1^2 +\partial_2^2)\sigma$.
We can define the complex coordinates $z = x^1 + i x^2$, $\bar{z} = x^1-ix^2$ and the derivatives $\partial_z = \frac{1}{2} (\partial_1 -i \partial_2)$, $\partial_{\bar{z}} = \frac{1}{2}( \partial_1+i \partial_2)$. The line element in these coordinates is
\be
ds^2 = e^{2 \sigma} dz d\bar{z},
\ee
and the Ricci scalar is $R = -8 e^{-2\sigma} \partial_z \partial_{\bar{z}}\sigma$.

\subsection{Symmetric tensors}
Given a tensor component in complex coordinates,  we can always trade it for a tensor component with only $z$ indices at the expense of introducing factors of $\hat{g}_{z \bar{z}}$ and $\hat{g}^{z \bar{z}}$. This means that we can restrict to tensors with only $z$ indices.
The helicity of a tensor is then defined as the number of upstairs $z$ indices minus the number of downstairs $z$ indices.   Define the one-dimensional complex vector bundle $\mathcal{T}^s$  as the space of all smooth helicity-$s$ tensors for $s \in \mathbb{Z}$. We can define an inner product on $\mathcal{T}^s$ as 
\be \label{eq:inner-product}
\langle T, T' \rangle = \intM dV (\hat{g}_{z \bar{z}})^s (T^{z \dots z})^* T'^{z \dots z},
\ee
where $dV$ denotes the Riemannian volume form of  $(\mathcal{M}, \hat{g})$. For simplicity, we write $\langle \cdot , \cdot \rangle$ rather than $\langle \cdot , \cdot \rangle_s$, since the helicity of the arguments of the inner product should be clear whenever this is important. The corresponding norm is denoted by $\norm{ \cdot}$.

Define $\mathcal{S}^s$ as the space of real, symmetric, traceless rank-$s$  tensors on $(\mathcal{M}, \hat{g})$ with $s\geq 1$. The nonvanishing components of a symmetric traceless tensor $T \in \mathcal{S}^s$  are 
\begin{align}
T_{z \dots z} & = \frac{1}{2} \left(T_{1 \dots 1} -i T_{1 \dots 1 2} \right), \\
T_{\bar{z} \dots \bar{z}}  = (\hat{g}_{z \bar{z}} )^sT^{z \dots z} &= \frac{1}{2} \left(T_{1 \dots 1} +i T_{1 \dots 1 2} \right),
\end{align}
and these components are related by complex conjugation, $T_{z \dots z} = (T_{\bar{z} \dots \bar{z}})^*$.
Thus the complexification of $\mathcal{S}^s$ is $\mathcal{T}^s \oplus \mathcal{T}^{-s}$ and the embedding of $\mathcal{S}^s$ in $\mathcal{T}^s \oplus \mathcal{T}^{-s}$ is given by 
\be
T_{p_1 p_2 \dots p_s} \mapsto 
\begin{pmatrix}
(\hat{g}^{z \bar{z}})^s (T_{z \dots z} )^*\\
T_{z \dots z} 
\end{pmatrix}.
\ee
Under this embedding, the canonical inner product on real symmetric tensors reduces to twice the real part of the inner product \eqref{eq:inner-product},
\begin{align}
\left( T, T' \right) & \coloneqq \intM dV \, T^{p_1 \dots p_s} T'_{p_1 \dots p_s} = \intM dV \, (\hat{g}_{z \bar{z}})^s \left[(T^{z \dots z})^* T'^{z \dots z} +T^{z \dots z} (T'^{z \dots z})^*\right] \\
& =  2  \, {\rm Re} \, \langle T, T' \rangle , \quad s\geq 1.
\end{align}

\subsection{Covariant derivatives and holomorphic $s$-differentials}
We can define two covariant derivatives acting on each space of helicity tensors \cite{Alvarez:1982zi}
\begin{align}
& \nabla_s^z: \mathcal{T}^s \rightarrow \mathcal{T}^{s+1}, \quad T^{z  \dots z} \mapsto \nabla_s^z T^{z  \dots z} = \hat{g}^{z \bar{z}} \partial_{\bar{z}} T^{z  \dots z},\\
& \nabla^s_z: \mathcal{T}^s \rightarrow \mathcal{T}^{s-1}, \quad T^{z  \dots z} \mapsto \nabla^s_z T^{z  \dots z} = (\hat{g}^{z \bar{z}})^s \partial_{z} \left[ (\hat{g}_{z \bar{z}})^sT^{z  \dots z} \right],
\end{align}
which agree with the components of the ordinary covariant derivative in complex coordinates. The formal adjoint of $\nabla_s^z$ with respect to the inner product in Eq.~\eqref{eq:inner-product} is $(\nabla_s^z)^{\dagger}=-\nabla_z^{s+1}$.
To simplify notation, we will often drop the helicity labels on the covariant derivatives when it is clear what bundles they act on. We also write $\nabla$ and $\bar{\nabla}$ for the coordinate-free expressions of $\nabla_z$ and $\nabla^z$, respectively.  A useful formula for the commutator of two covariant derivatives is
\begin{align} \label{eq:nabla-commutator}
[\nabla^z, \nabla_z]: \mathcal{T}^s \rightarrow \mathcal{T}^{s}, \quad T^{z  \dots z} \mapsto s T^{z  \dots z} .
\end{align}

Let us mention the connection between symmetric, transverse, traceless tensors and holomorphic $s$-differentials. This connects the approach of the present paper to that of Ref.~\cite{Bonifacio:2021msa}. A real, symmetric, traceless tensor $T \in\mathcal{S}^s$ is transverse if $\nabla^{p_1} T_{p_1 \dots p_s}=0$. For $s \geq 2$, the helicity components of such a tensor satisfy
\be
 \nabla^z T_{z z \dots z} =\nabla^{\bar{z}} T_{\bar{z} \bar{z} \dots \bar{z}} =0  \implies \partial_z T_{\bar{z} \dots \bar{z}} =\partial_{\bar{z}} T_{z \dots z} =0,
\ee
which implies that $T_{z \dots z} $ is holomorphic and $T_{\bar{z} \dots \bar{z}} $ is anti-holomorphic. A holomorphic tensor $T_{z \dots z} \in \mathcal{T}^{-s}$ is called a holomorphic $s$-differential (or a holomorphic differential when $s=1$ and a holomorphic quadratic differential when $s=2$). The Riemann--Roch theorem implies that the vector space of holomorphic $s$-differentials with $s \geq 2$ has complex dimension $(2s-1)(g-1)$ on a closed surface of genus $g \geq 2$.
The helicity components of a real transverse vector satisfy $\nabla_z T^z + \nabla^{z} T_{z} =0$. This does not imply that $T_z$ is a holomorphic differential since, in addition to the holomorphic part, such a vector can have a longitudinal component $i \partial_z \phi$. On a closed surface of genus $g$, the complex vector space of holomorphic differentials has dimension $g$.

\subsection{Tensor decompositions}
We now discuss some orthogonal decompositions for helicity tensors. Consider the covariant derivative $\nabla_z^{-s+1}: \mathcal{T}^{-s+1} \rightarrow \mathcal{T}^{-s}$ for $s \geq 1$. It has  adjoint $(\nabla_z^{-s+1})^{\dagger} = - \nabla^z_{-s}$, whose kernel consists of holomorphic $s$-differentials. Since $\nabla_z^{-s+1}$ is a linear elliptic operator, we have the orthogonal decomposition
\be
\mathcal{T}^{-s} = \ker ((\nabla_z^{-s+1})^{\dagger}  )\oplus {\rm Im}(\nabla_z^{-s+1})= \ker (\nabla^z_{-s}) \oplus {\rm Im}(\nabla_z^{-s+1}).
\ee
This means that we can write any $T^{(-s)} \in \mathcal{T}^{-s}$ as
\be
T^{(-s)}_{z \dots z} = \tilde{T}^{(-s)}_{z \dots z} + \nabla^{-s+1}_z T^{(-s+1)}_{z \dots z},
\ee
where $\tilde{T}^{(-s)}$ is a holomorphic $s$-differential and $T^{(-s+1)} \in \mathcal{T}^{-s+1}$.
Applying this repeatedly gives the orthogonal decomposition 
\be \label{eq:-s-decomposition}
T^{(-s)}_{z \dots z} = \sum_{j=0}^{s-1} (\nabla_z)^{j} \tilde{T}^{(-s+j)}_{z \dots z}+ (\nabla_z)^s T^{(0)},
\ee
where $\tilde{T}^{(-k)}$ is a holomorphic $k$-differential, $T^{(0)} \in \mathcal{T}^0$ is a smooth function, and $(\nabla_z)^{j}$ stands for $\nabla_z$ applied $j$ times.\footnote{This should not be confused with $\nabla_z^j$, which is one of the covariant derivatives acting on $\mathcal{T}^{j}$. The expanded form of Eq.~\eqref{eq:-s-decomposition} is
\begin{align}
T^{(-s)}_{z \dots z} =&  \tilde{T}^{(-s)}_{z \dots z} + \nabla^{-s+1}_z  \tilde{T}^{(-s+1)}_{z \dots z} +\dots +\nabla^{-s+1}_z \nabla^{-s+2}_z \dots \nabla^{-1}_z \tilde{T}^{(-1)}_z +\nabla^{-s+1}_z \nabla^{-s+2}_z \dots \nabla^{0}_z T^{(0)}.
\end{align}
}

Similarly, for $s \geq 1$ the linear elliptic operator $\nabla_{s-1}^z: \mathcal{T}^{s-1} \rightarrow \mathcal{T}^s$ has the formal adjoint $-\nabla_z^s$, whose kernel consists of  tensors $T^{z \dots z}$ such that $T_{\bar{z} \dots \bar{z}} = (\hat{g}_{z \bar{z}})^s T^{z \dots z}$ is antiholomorphic. From the orthogonal decomposition
\be
\mathcal{T}^{s} = \ker ((\nabla_{s-1}^z)^{\dagger}  )\oplus {\rm Im}(\nabla_{s-1}^z)= \ker (\nabla_z^s) \oplus {\rm Im}(\nabla^z_{s-1}),
\ee
we can write any $T^{(s)} \in \mathcal{T}^s$ as
\be
T^{(s) z \dots z} = \tilde{T}^{(s) z \dots z} + \nabla^z_{s-1} T^{(s-1) z \dots z},
\ee
where $\tilde{T}^{(s)} $ is in the kernel of $\nabla_z^s$ and $T^{(s-1)} \in \mathcal{T}^{s-1}$. Applying this repeatedly gives the orthogonal decomposition
\be \label{eq:+s-decomposition}
T^{(s) z \dots z} = \sum_{j=0}^{s-1} (\nabla^z)^j \tilde{T}^{(s-j) z \dots z} + (\nabla^z)^s T^{(0) },
\ee
where $\tilde{T}^{(k)} $ is in the kernel of $\nabla_z^k$ and $T^{(0)} \in \mathcal{T}^0$.

\subsection{Laplacians}
We can define two Laplacians on helicity tensors,
\be
\Delta^{(+)} \coloneqq - 2\nabla_z \nabla^z , \quad \Delta^{(-)} \coloneqq - 2\nabla^z \nabla_z ,
\ee
where we leave implicit the helicity of the tensors on which these operators act. When acting on $\mathcal{T}^s$,  Eq.~\eqref{eq:nabla-commutator} gives
\be \label{eq:delta-diff}
\Delta^{(+)}-\Delta^{(-)} = 2s.
\ee

Let us see how these Laplacians act on the terms appearing in the tensor decompositions considered above. Firstly, we can decompose any $L^2$-normalisable function in terms of eigenfunctions of the Laplace--Beltrami operator $\Delta$. 
For an eigenfunction $\phi_i$ of $\Delta$ with eigenvalue $\lambda_i$, we have 
\be
\Delta \phi_i = \Delta^{(\pm)} \phi_i = \lambda_i \phi_i. 
\ee
This tells us how the Laplacians $\Delta^{(\pm)}$ act on the scalar components of the tensor decompositions. We take the eigenfunctions $\left\{\phi_i\right\}_{i=0}^{\infty}$ to be real and orthonormal, with the normalisation $\left\langle \phi_i, \phi_j \right\rangle =\delta_{ij}$.

Now let $\left\{ \phi^{(s)}_{i} \right\}_{i=1}^{N_{g,s}}$ with $s \geq 1$ be a finite-dimensional orthonormal basis  of holomorphic $s$-differentials, with the normalisation $\left\langle \phi_i^{(s)}, \phi_j^{(s)} \right\rangle=\delta_{ij}$.\footnote{Note that we use $s$ rather than $-s$ in the superscript for the basis of holomorphic $s$-differentials.}  As already mentioned, on a genus-$g$ surface with $g\geq 2$, the Riemann--Roch theorem gives the complex dimension of the vector space of holomorphic $s$-differentials as
\be
N_{g,s } \coloneqq 
\begin{cases*}
g& if  $s =1$  \\
(2s-1)(g-1)  & if $s >1$
\end{cases*}.
\ee
We can write any holomorphic $s$-differential $\tilde{T}^{(-s)}\in\mathcal{T}^{-s}$ as a complex linear combination of these $\phi^{(s)}_{i}$. Using Eq.~\eqref{eq:delta-diff} and the fact that $\phi^{(s)}_{i}$ is in the kernel of $\nabla^z$, we get
\be \label{eq:s-eigen-equation}
\Delta^{(+)} \phi^{(s)}_{i, z \dots z} =0, \quad \Delta^{(-)} \phi^{(s)}_{i, z \dots z}= 2s \phi^{(s)}_{i, z \dots z},
\ee
i.e., holomorphic $s$-differentials are eigentensors of $\Delta^{(\pm)}$ with eigenvalues $\lambda_{i, s}^{(\pm)} = s \mp s$. 

Now let us define
\be
\bar{\phi}^{(s)z \dots z}_{i} \coloneqq (\hat{g}^{z \bar{z}})^s \left(\phi^{(s)}_{i, z \dots z}\right)^*.
\ee
Any $\tilde{T}^{(s)} \in\mathcal{T}^{s}$ in the kernel of $\nabla_z^s$ can be written as a complex linear combination of these $\bar{\phi}^{(s)}_{i}$. Using Eq.~\eqref{eq:delta-diff} and the fact that $\bar{\phi}^{(s)}_{i} $ is in the kernel of $\nabla_z$, we get
\be \label{eq:-s-eigen-equation}
\Delta^{(-)} \bar{\phi}^{(s)z \dots z}_{i}   = 0, \quad \Delta^{(+)} \bar{\phi}^{(s)z \dots z}_{i}   = 2s\bar{\phi}^{(s)z \dots z}_{i} ,
\ee
i.e., tensors in the kernel of $\nabla_z^s$ are eigentensors of $\Delta^{(\pm)}$ with eigenvalues $\bar{\lambda}_{i, s}^{(\pm)}=s\pm s$. 

It is convenient to define $\phi_i^{(0)} \coloneqq \phi_i$ and $\bar{\phi}_i^{(0)} \coloneqq \phi_i$ for $i=1, 2, \dots$, even though these eigenfunctions are not holomorphic functions. It is similarly convenient to define $N_{g, 0} \coloneqq \infty$, $\lambda_{i, 0}^{(\pm)} \coloneqq \lambda_i$, and $\bar{\lambda}_{i, 0}^{(\pm)} \coloneqq \lambda_i$.  We can then write $\phi^{(s)}_i$ with $s \geq 0$ to collectively denote nonconstant scalar eigenfunctions and holomorphic $s$-differentials, where $1 \leq i \leq N_{g, s}$, and similarly for $\bar{\phi}^{(s)}_i$. From now on, when we write $\phi^{(s)}_i$ or $\bar{\phi}^{(s)}_i$, this can include the scalar eigenfunctions, unless explicitly indicated otherwise. We also now take $\phi_i$ to have $i \geq 1$ unless indicated otherwise, since the zero mode $\phi_0 = V^{-1/2}$ will be treated separately.

\subsection{Spectral decompositions}
We can combine the helicity tensor decomposition from Eq.~\eqref{eq:-s-decomposition} with the decompositions of functions in terms of $ \left\{ \phi_i \right\}_{i=0}^{\infty}$ and holomorphic $s$-differentials in terms of $\left\{ \phi^{(s>0)}_{i} \right\}_{i=1}^{N_{g,s}}$ to write a tensor $T^{(-s)}\in\mathcal{T}^{-s}$ with nonpositive helicity as
\be \label{eq:-s-eigen-decomposition}
T^{(-s)}_{z \dots z} = \sum_{n=0}^{s} \sum^{N_{g,n}}_{k=1} C_{k, n} (\nabla_z)^{s-n} \phi^{(n)}_{k, z \dots z} + \delta_{s 0}C_{0,0}V^{-\frac{1}{2}} ,
\ee
where $C_{k,n}$ are constants that can be determined by taking inner products of each side of this equation with the pairwise orthogonal terms on the right-hand side. Similarly, using the helicity tensor decomposition from Eq.~\eqref{eq:+s-decomposition}, we can write a tensor $T^{(s) }\in\mathcal{T}^{s}$ with nonnegative helicity  as
\be \label{eq:+s-eigen-decomposition}
T^{(s) z \dots z} = \sum_{n=0}^{s} \sum^{N_{g,n}}_{k=1} \bar{C}_{k, n} (\nabla^z)^{s-n} \bar{\phi}^{(n) z \dots z}_{k} + \delta_{s 0}\bar{C}_{0,0}V^{-\frac{1}{2}} ,
\ee
where $\bar{C}_{k,n}$ are constants that can be determined by taking inner products of each side of this equation with the pairwise orthogonal terms on the right-hand side. 

\subsection{Triple overlap integrals}
We now define certain basic integrals of products of eigenmodes. 
Given three eigenmodes $\phi_i^{(s_1)}$, $\phi_j^{(s_2)}$, and $\phi_k^{(s_3)}$, with $s_3 \geq s_1+s_2$, we define the following triple overlap integral:
\be
c^{(s_1,s_2,s_3)}_{ijk} \coloneqq  \left\langle \phi_k^{(s_3)}, \phi_i^{(s_1)} \nabla^{s_3-s_1-s_2} \phi_j^{(s_2)} \right\rangle= \int_{\mathcal{M}} dV \phi^{(s_1)}_{i, z\dots z} (\nabla_z)^{s_3-s_1-s_2} \phi^{(s_2)}_{j, z\dots z}  \bar{\phi}_k^{(s_3) z\dots z},
\ee
where $\nabla$ is the coordinate-free expression for $\nabla_z$. 
The complex conjugate of $c^{(s_1,s_2,s_3)}_{ijk} $ is denoted by $\bar{c}^{(s_1,s_2,s_3)}_{ijk}$, so for $s_3 \geq s_1+s_2$ we have
\be
\bar{c}^{(s_1,s_2,s_3)}_{ijk} =\left\langle \phi_i^{(s_1)} \nabla^{s_3-s_1-s_2} \phi_j^{(s_2)}, \phi_k^{(s_3)} \right\rangle= \int_{\mathcal{M}} dV \bar{\phi}^{(s_1)z\dots z}_{i} (\nabla^z)^{s_3-s_1-s_2} \bar{\phi}^{(s_2)z\dots z}_{j}  \phi_{k, z\dots z}^{(s_3)}.
\ee

These triple overlap integrals satisfy various relations. For example, we have 
\be
c^{(s_1,s_2,s_3)}_{ijk}=(-1)^{s_3-s_1-s_2}c^{(s_2,s_1,s_3)}_{jik},
\ee
which implies in particular that $c_{iij}^{(s_1, s_1, s_3)} $ vanishes when $s_3$ is odd. The triple overlap integrals are complex in general, but $c_{ijk}^{(0, s, s)}$ satisfies
\be
c_{ijk}^{(0, s, s)} = \bar{c}_{ikj}^{(0, s, s)},
\ee
which implies that $c_{ijj}^{(0, s, s)}$ is real. Bounds on the asymptotic growth of the scalar overlap integrals $c_{iik}^{(0,0,0)}$ as $k \rightarrow \infty$ are known from analytic number theory \cite{Sarnak94, Petridis95, Bernstein1999}. Triple overlap integrals can also be constrained using bootstrap methods \cite{Bonifacio:2020xoc, Bonifacio:2021msa}, as with the conformal bootstrap bounds on operator product expansion coefficients \cite{Caracciolo:2009bx}.

The overlap integrals written so far do not exhaust all possibilities. There is an additional independent triple overlap integral between three distinct scalar eigenfunctions that we write as
\be
\tilde{c}_{ijk}\coloneqq \frac{\sqrt{-1}}{2} \left( \left\langle \nabla \phi_k, \phi_{i} \nabla  \phi_{j} \right\rangle-\left\langle \nabla \phi_j, \phi_{i} \nabla  \phi_{k} \right\rangle \right) = \sqrt{-1}\intM dV \phi_{i} \nabla_z \phi_{[j}  \nabla^z \phi_{k]},
\ee
where we antisymmetrise with weight one.
It is real and completely antisymmetric in its indices,
\be
\bar{\tilde{c}}_{ijk} = \tilde{c}_{ijk}, \quad \tilde{c}_{ijk}=\tilde{c}_{[ijk]} .
\ee
 This overlap integral corresponds to the parity-odd interaction $ \epsilon^{mn} \phi_i \partial_{m} \phi_j \partial_n \phi_k$. With this definition, we can write
 \be \label{eq:011-identity}
  \left\langle \nabla \phi_k, \phi_{i} \nabla  \phi_{j} \right\rangle = -\sqrt{-1}\tilde{c}_{ijk}+ \frac{1}{4}\left( \lambda_j+\lambda_k-\lambda_i\right) c_{ijk}^{(0,0,0)}.
 \ee

\section{Consistency conditions}
\label{sec:ccs}

In this section, we show how to write down consistency conditions involving triple overlap integrals and Laplacian eigenvalues. We start by discussing  integrals of products of two eigenmodes and work up to integrals of products of four eigenmodes.

\subsection{Norms}
Consider first the integral of a product of derivatives of two eigenmodes of the same rank,
\be
\left\langle \nabla^{\alpha} \phi^{(s)}_j , \nabla^{\alpha} \phi^{(s)}_i \right\rangle = \int_{\mathcal{M}} dV (\nabla_z)^{\alpha} \phi^{(s)}_{i, z \dots z} (\nabla^z)^{\alpha} \bar{\phi}^{(s) z \dots z}_{j} ,
\ee
where $\alpha$ and $s$ are nonnegative integers. We can simplify this using integration by parts and the normalisations of the eigenmodes. 
A useful identity for this is
\begin{align}
\nabla^z (\nabla_z)^{\alpha} \phi_{i, z \dots z}^{(s)} & = -\frac{1}{2}\left( \alpha(2s+\alpha-1) + \lambda_{i, s}^{(+)} \right) (\nabla_z)^{\alpha-1}  \phi^{(s)}_{i, z \dots z} , 
\end{align}
where $\alpha \geq 1$ and we recall that $\lambda_{i, s>0}^{(+)} =0$ and $\lambda_{i, 0}^{(+)} = \lambda_i$.
This identity can be derived by repeatedly using Eq.~\eqref{eq:nabla-commutator}.
We can use this identity to recursively solve for the norms. For $s\geq 1$, we can write the result as
\be
\left\langle \nabla^{\alpha} \phi^{(s)}_j , \nabla^{\alpha} \phi^{(s)}_i \right\rangle =\frac{1}{2^{\alpha}} \frac{ \Gamma(\alpha+1)\Gamma(2s+\alpha)}{\Gamma(2s)} \delta_{ij},
\ee
and for $s=0$ we can write
\be \label{eq:scalar-norm}
\left\langle \nabla^{\alpha} \phi_j , \nabla^{\alpha} \phi_i \right\rangle = \delta_{ij} \chi_{\alpha}(\lambda_i), \quad \chi_{\alpha}(\lambda_i) \coloneqq \frac{1}{2^{\alpha}} \prod_{m=0}^{\alpha-1} \left[\lambda_i + m(m+1)\right].
\ee

\subsection{Triple products}
We now consider integrals of products of derivatives of three eigenmodes of the form
\be \label{eq:gen-cubic}
\left\langle \nabla^{\alpha_3} \phi^{(s_3)}_k,  \nabla^{\alpha_1} \phi^{(s_1)}_i \nabla^{\alpha_2} \phi^{(s_2)}_j \right\rangle = \int_{\mathcal{M}} dV \, (\nabla_z)^{\alpha_1} \phi^{(s_1)}_{i, z\dots z} (\nabla_z)^{\alpha_2} \phi^{(s_2)}_{j, z \dots z}   (\nabla^z)^{\alpha_3} \bar{\phi}^{(s_3) z \dots z}_k ,
\ee
where $\alpha_a$ and $s_a$ are nonnegative integers satisfying
\be
\alpha_1 + \alpha_2 +s_1+s_2=\alpha_3+s_3.
\ee
Using integration by parts, we can write Eq.~\eqref{eq:gen-cubic} in terms of the basic triple overlap integrals defined earlier. A useful identity for doing this, which is valid when $\alpha_1+s_1>0$, $\alpha_2+s_2>0$, and $\alpha_3>0$, is
\begin{align}
 & \left\langle \nabla^{\alpha_3} \phi^{(s_3)}_k,  \nabla^{\alpha_1} \phi^{(s_1)}_i \nabla^{\alpha_2} \phi^{(s_2)}_j \right\rangle  = \frac{(1-\delta_{\alpha_1 0})}{2}\left( \alpha_1 (2 s_1+\alpha_1 -1) + \lambda_{i, s_1}^{(+)}\right) \left\langle \nabla^{\alpha_3-1} \phi^{(s_3)}_k,  \nabla^{\alpha_1-1} \phi^{(s_1)}_i \nabla^{\alpha_2} \phi^{(s_2)}_j \right\rangle  \nn \\
& + \frac{(1-\delta_{\alpha_2 0})}{2}\left( \alpha_2 (2 s_2+\alpha_2 -1) + \lambda_{j, s_2}^{(+)}\right) \left\langle \nabla^{\alpha_3-1} \phi^{(s_3)}_k,  \nabla^{\alpha_1} \phi^{(s_1)}_i \nabla^{\alpha_2-1} \phi^{(s_2)}_j \right\rangle.
\end{align}
A useful identity when $s_1=\alpha_1=0$, which is valid for $\alpha_2>1$ and $\alpha_3 +s_3 >1$, is
{\fontsize{10}{10}\selectfont 
\begin{align}
 &\hspace{-1.5cm} \left\langle \nabla^{\alpha_3} \phi^{(s_3)}_k,  \phi_i \nabla^{\alpha_2} \phi^{(s_2)}_j \right\rangle  =   \frac{1}{2}\left( \alpha_2 (2 s_2+\alpha_2 -1)+(\alpha_3-1)(2s_3+\alpha_3-2) + \lambda_{j, s_2}^{(+)}+\bar{\lambda}_{k, s_3}^{(-)}- \lambda_{i}\right)  \left\langle \nabla^{\alpha_3-1} \phi^{(s_3)}_k, \phi_i \nabla^{\alpha_2-1} \phi^{(s_2)}_j \right\rangle  \nn \\
& \hspace{-1.35cm}- \frac{(1-\delta_{\alpha_3 1})}{4}\left( ( \alpha_2-1) (2 s_2+\alpha_2 -2) + \lambda_{j, s_2}^{(+)}\right)\left( ( \alpha_3-1) (2 s_3+\alpha_3 -2) + \bar{\lambda}_{k, s_3}^{(-)}\right) \left\langle  \nabla^{\alpha_3-2} \phi^{(s_3)}_k,  \phi_i \nabla^{\alpha_2-2} \phi^{(s_2)}_j  \right\rangle,
\end{align}
}where we recall that $\bar{\lambda}_{i, s>0}^{(-)} =0$ and $\bar{\lambda}_{i, 0}^{(-)} = \lambda_i$.
By recursively using these identities and their complex conjugates, plus Eq.~\eqref{eq:011-identity}, we can write Eq.~\eqref{eq:gen-cubic} and its conjugate in terms of the eigenvalues $\lambda_{a}$ and the basic triple overlap integrals $c_{abc}^{(s_a, s_b, s_c)}$, $\bar{c}_{abc}^{(s_a, s_b, s_c)}$, and $\tilde{c}_{abc}$. 

\subsection{Quadruple products}
Now that we know how to reduce double and triple overlap integrals, we can explain how to obtain consistency conditions from quadruple overlap integrals. We consider integrals of products of derivatives of four eigenmodes taking the form
\be \label{eq:gen-quartic}
\left\langle \nabla^{\alpha_3} \phi^{(s_3)}_k \nabla^{\alpha_4} \phi^{(s_4)}_l,  \nabla^{\alpha_1} \phi^{(s_1)}_i \nabla^{\alpha_2} \phi^{(s_2)}_j \right\rangle = \int_{\mathcal{M}} dV \, (\nabla_z)^{\alpha_1} \phi^{(s_1)}_{i, z\dots z} (\nabla_z)^{\alpha_2} \phi^{(s_2)}_{j, z \dots z}  (\nabla^z)^{\alpha_3} \bar{\phi}^{(s_3) z \dots z}_k   (\nabla^z)^{\alpha_4} \bar{\phi}^{(s_4) z \dots z}_l, 
\ee
where $\alpha_a$ and $s_a$ are nonnegative integers satisfying
\be
J_s \coloneqq \alpha_1 + \alpha_2 +s_1+s_2=\alpha_3+\alpha_4+s_3+s_4.
\ee
The idea is to write this quadruple overlap integral in terms of triple overlap integrals in multiple ways using the helicity tensor decompositions discussed earlier. 

One way to rewrite the quadruple overlap integral is to first use Eq.~\eqref{eq:-s-eigen-decomposition} to expand the product $\nabla^{\alpha_1} \phi^{(s_1)}_{i} \nabla^{\alpha_2} \phi^{(s_2)}_{j}$ on a genus-$g$ surface as
\begin{align}
(\nabla_z)^{\alpha_1} \phi^{(s_1)}_{i, z \dots z} (\nabla_z)^{\alpha_2} \phi^{(s_2)}_{j, z \dots z} & = \sum_{n=0}^{J_s} \sum^{N_{g,n}}_{k'=1} \frac{\left\langle \nabla^{J_s-n} \phi^{(n)}_{k'},  \nabla^{\alpha_1} \phi^{(s_1)}_i \nabla^{\alpha_2} \phi^{(s_2)}_j \right\rangle }{ \norm{ \nabla^{J_s-n} \phi^{(n)}_{k'}}^2 } (\nabla_z)^{J_s-n} \phi^{(n)}_{k', z \dots z} \nn \\
& + \frac{\delta_{ij} \delta_{J_s, 0}}{V} .
\end{align}
We can similarly expand $ \bar{\nabla}^{\alpha_3} \bar{\phi}^{(s_3) }_k  \bar{\nabla}^{\alpha_4} \bar{\phi}^{(s_4)}_l$ using Eq.~\eqref{eq:+s-eigen-decomposition},
\begin{align}
(\nabla^z)^{\alpha_3} \bar{\phi}^{(s_3) z \dots z}_{k} (\nabla^z)^{\alpha_4} \bar{\phi}^{(s_4)z \dots z}_{l} & = \sum_{n=0}^{J_s} \sum^{N_{g,n}}_{k'=1} \frac{\left\langle   \nabla^{\alpha_3} \phi^{(s_3)}_k \nabla^{\alpha_4} \phi^{(s_4)}_l , \nabla^{J_s-n} \phi^{(n)}_{k'}\right\rangle }{ \norm{\nabla^{J_s-n} \phi^{(n)}_{k'} }^2} (\nabla^z)^{J_s-n} \bar{\phi}^{(n)  z \dots z}_{k'} \nn \\
& + \frac{\delta_{kl} \delta_{J_s, 0}}{V} .
\end{align}
Substituting these two expansions into Eq.~\eqref{eq:gen-quartic} then gives the decomposition of the quadruple overlap integral,
\begin{align} 
& \left\langle \nabla^{\alpha_3} \phi^{(s_3)}_k \nabla^{\alpha_4} \phi^{(s_4)}_l,  \nabla^{\alpha_1} \phi^{(s_1)}_i \nabla^{\alpha_2} \phi^{(s_2)}_j \right\rangle  = \frac{ \delta_{ij} \delta_{kl} \delta_{J_s 0}}{V} \nn \\
& +\sum_{n=0}^{J_s} \sum^{N_{g,n}}_{k'=1} \frac{\left\langle   \nabla^{\alpha_3} \phi^{(s_3)}_k \nabla^{\alpha_4} \phi^{(s_4)}_l , \nabla^{J_s-n} \phi^{(n)}_{k'}\right\rangle \left\langle \nabla^{J_s-n} \phi^{(n)}_{k'},  \nabla^{\alpha_1} \phi^{(s_1)}_i \nabla^{\alpha_2} \phi^{(s_2)}_j \right\rangle}{ \norm{\nabla^{J_s-n} \phi^{(n)}_{k'} }^2}.
\end{align}
In physics terminology, this is called the $s$-channel decomposition of the quadruple overlap integral. Alternatively, we could first write the same quadruple overlap integral as
\be
\left\langle  \bar{\nabla}^{\alpha_2} \bar{\phi}^{(s_2)}_j \nabla^{\alpha_4} \phi^{(s_4)}_l,  \nabla^{\alpha_1} \phi^{(s_1)}_i  \bar{\nabla}^{\alpha_3} \bar{\phi}^{(s_3)}_k  \right\rangle  
\ee
and then expand the two products using the helicity tensor decompositions. This gives the $t$-channel decomposition of the integral,
\begin{align}
& \left\langle  \bar{\nabla}^{\alpha_2} \bar{\phi}^{(s_2)}_j \nabla^{\alpha_4} \phi^{(s_4)}_l,  \nabla^{\alpha_1} \phi^{(s_1)}_i  \bar{\nabla}^{\alpha_3} \bar{\phi}^{(s_3)}_k  \right\rangle  = \frac{ \left\langle  \nabla^{\alpha_3} \phi^{(s_3)}_k , \nabla^{\alpha_1} \phi^{(s_1)}_{i} \right\rangle  \left\langle  \nabla^{\alpha_4} \phi^{(s_4)}_l , \nabla^{\alpha_2} \phi^{(s_2)}_{j} \right\rangle \delta_{J_t, 0}}{V} \nn \\
& +\sum_{n=0}^{J_t} \sum^{N_{g,s}}_{k'=1} \frac{\left\langle  \bar{\nabla}^{\alpha_2} \bar{\phi}^{(s_2)}_j \nabla^{\alpha_4} \phi^{(s_4)}_l , \nabla^{J_t-n} \phi^{(n)}_{k'}\right\rangle \left\langle \nabla^{J_t-n} \phi^{(n)}_{k'},  \nabla^{\alpha_1} \phi^{(s_1)}_i  \bar{\nabla}^{\alpha_3} \bar{\phi}^{(s_3)}_k  \right\rangle}{\norm{ \nabla^{J_t-n} \phi^{(n)}_{k'}}^2 },
\end{align}
where $J_t \coloneqq |s_1 +\alpha_1 -s_3-\alpha_3|$ and we have written the expansion assuming that $s_1 +\alpha_1 \geq s_3 +\alpha_3$.
Finally, we can exchange $3\leftrightarrow 4$ in the $t$-channel decomposition to get the $u$-channel decomposition.

As mentioned previously, we can write the inner products appearing in these decompositions in terms of the eigenvalues $\lambda_a$ and the triple overlap integrals  $c_{abc}^{(s_a, s_b, s_c)}$, $\bar{c}_{abc}^{(s_a, s_b, s_c)}$, and $\tilde{c}_{abc}$. Equating the three decompositions, we then obtain two consistency conditions involving just eigenvalues and the basic triple overlap integrals. 
As an example, for the $s$- and $t$-channel decompositions of $\left\langle \phi_i \nabla \phi^{(1)}_j,   \phi_i \nabla\phi^{(1)}_j \right\rangle$ we have
\begin{align}
\left\langle \phi_i \nabla \phi^{(1)}_j,   \phi_i \nabla\phi^{(1)}_j \right\rangle & =  \sum_{k=1}^{\infty}\frac{ (\lambda_k-\lambda_i+2)^2 }{\lambda_k(2+\lambda_k)}\left| c^{(0,0,1)}_{i k j}\right|^2+\frac{1}{4} \sum_{k=1}^{N_{g,1}}(\lambda_i-2)^2 \left| c^{(0,1,1)}_{ij k}\right|^2+  \sum_{k=1}^{N_{g,2}}\left| c^{(0,1,2)}_{ij k}\right|^2, \\
\left\langle  \bar{\nabla} \bar{\phi}^{(1)}_j \nabla \phi^{(1)}_j,   \phi_i  \phi_i \right\rangle & = \frac{1}{V} + \sum_{k=1}^{\infty}\left( 1- \frac{\lambda_k}{2} \right) c^{(0,0,0)}_{iik} c^{(0,1,1)}_{kjj},
\end{align}
and equating these expressions gives a consistency condition.

\section{Eigenvalue bounds}
\label{sec:bounds}

In this section, we illustrate how to use consistency conditions to derive bounds on the eigenvalues of closed hyperbolic surfaces. This approach was used to derive bounds for closed hyperbolic manifolds in general dimensions in Ref.~\cite{Bonifacio:2021msa}, using consistency conditions coming from quadruple overlap integrals of a fixed scalar eigenfunction with up to 16 derivatives (and with up to six derivatives for closed Einstein manifolds in Ref.~\cite{Bonifacio:2020xoc}).  The approach closely follows that of the conformal bootstrap \cite{Rattazzi:2008pe, Rychkov:2009ij, Poland:2018epd}. To derive bounds, we use the arbitrary-precision semidefinite program solver \texttt{SDPB}  \cite{Simmons-Duffin:2015qma, Landry:2019qug}. We have not computed rigorous errors for our bounds, although we expect that they are correct to the stated precision and  that they could be made rigorous with additional effort.

\subsection{Bootstrap equations}

To derive bootstrap bounds, we consider real quadruple overlap integrals involving at most two different eigenfunctions or holomorphic $s$-differentials and taking the form
\be \label{eq:special-quartic}
\left\langle \nabla^{\alpha} \phi^{(s_1)}_i \phi^{(s_2)}_j,  \nabla^{\alpha} \phi^{(s_1)}_i  \phi^{(s_2)}_j \right\rangle = \int_{\mathcal{M}} dV \, (\nabla_z)^{\alpha} \phi^{(s_1)}_{i, z\dots z}  \phi^{(s_2)}_{j, z \dots z}  (\nabla^z)^{\alpha} \bar{\phi}^{(s_1) z \dots z}_i   \bar{\phi}^{(s_2) z \dots z}_j \, ,
\ee
where $\alpha$, $s_1$, and $s_2$ are nonnegative  integers with $s_1 \leq s_2$. We also assume that $s_2>0$ to avoid the additional complications present for two distinct fixed scalar eigenfunctions, but similar formulae can be written down for this case.

We can be reasonably explicit in writing down the different decompositions of Eq.~\eqref{eq:special-quartic}. On a surface of genus $g$, we have the $s$-channel decomposition 
\begin{align}
&\hspace{-1cm} \left\langle \nabla^{\alpha} \phi^{(s_1)}_i \phi^{(s_2)}_j,  \nabla^{\alpha} \phi^{(s_1)}_i  \phi^{(s_2)}_j \right\rangle = \frac{\delta_{ij} \delta_{J_s 0}}{V}+ \sum_{n=s_1+s_2}^{J_s} \sum_{k=1}^{N_{g,n}}  \frac{ \Gamma(2n)\left[\prod_{m=1}^{J_s-n}\left( (\alpha+1-m)(2s_1+\alpha-m) +\lambda_{i,n}^{(+)} \right)\right]^2\left| c_{ijk}^{(s_1,s_2, n)}\right|^2}{ 2^{J_s-n}\Gamma(J_s+n) \Gamma(J_s-n+1)} \nn \\
& \hspace{-1cm}+ \delta_{s_1 0}\left[ \sum_{k=1}^{\infty}  \frac{ \left[\chi_{\alpha}(\lambda_i)\right]^2 \left| c_{kij}^{(0, 0, s_2)}\right|^2}{ 2^{s_1}\chi_{J_s}(\lambda_k) }+  \sum_{n=1}^{s_2-1} \sum_{k=1}^{N_{g,n}}  \frac{2^{\alpha+s_2-n} \Gamma(2n) \left[\chi_{\alpha}(\lambda_i) \right]^2\left| c_{kij}^{(n,0, s_2)}\right|^2}{ \Gamma(J_s+n)  \Gamma(J_s-n+1)} \right],
\end{align}
the $t$-channel decomposition 
\be
\left\langle \phi^{(s_2)}_j  \bar{\phi}^{(s_2)}_j ,  \nabla^{\alpha} \phi^{(s_1)}_i \bar{\nabla}^{\alpha} \bar{\phi}^{(s_1)}_i \right\rangle = \frac{1}{V}\norm{\nabla^{\alpha} \phi^{(s_1)}_i}^2+ \sum_{k=1}^{\infty} f_{s_1}^{( \alpha)} \!\left( \lambda_{i, s}^{(+)}, \lambda_k \right) c_{kii}^{(0,s_1,s_1)}c_{kjj}^{(0,s_2,s_2)},
\ee    
and the $u$-channel decomposition 
\be
\left\langle \nabla^{\alpha} \phi^{(s_1)}_i \bar{\phi}^{(s_2)}_j,  \nabla^{\alpha} \phi^{(s_1)}_i  \bar{\phi}^{(s_2)}_j \right\rangle = 
\frac{\delta_{ij} \delta_{J_u 0} \delta_{\alpha 0}}{V}+ \sum_{k=1}^{\infty}  \chi_{J_u}(\lambda_k)  \left| c_{kij}^{(0, s_1,s_2)}\right|^2 
\ee
if $s_1+\alpha \geq s_2$, or 
\be
\left\langle \nabla^{\alpha} \phi^{(s_1)}_i \bar{\phi}^{(s_2)}_j,  \nabla^{\alpha} \phi^{(s_1)}_i  \bar{\phi}^{(s_2)}_j \right\rangle = 
\sum_{k=1}^{\infty}  \frac{\left| c_{kij}^{( 0, s_1 ,s_2)}\right|^2}{  \chi_{J_u}(\lambda_k) } +\sum_{n=1}^{J_u} \sum_{k=1}^{N_{g,n}}  \frac{2^{J_u-n} \Gamma(2n)\left| c_{kij}^{( n , s_1, s_2)}\right|^2}{ \Gamma(J_u+n) \Gamma(J_u-n+1)} 
\ee
if $s_1+\alpha < s_2$,
where $J_s \coloneqq s_1+s_2+\alpha$, $J_u \coloneqq |\alpha+ s_1-s_2|$, $\chi_{\alpha}(\lambda)$ are the polynomials defined in Eq.~\eqref{eq:scalar-norm}, and $f_{s}^{(\alpha)}\!\left( \lambda_{i, s}^{(+)}, \lambda_k \right)$ is the solution to the recursion relation
\begin{align}
 f_{s}^{( \alpha) }& = \frac{1}{2}\left( \alpha(2 s+\alpha-1)+(\alpha-1)(2s+\alpha-2)+2\lambda_{i, s}^{(+)}-\lambda_{k} \right) f_{s}^{(\alpha-1)} \nn \\
 & -\frac{1}{4}\left[ \left(\alpha-1\right) \left( 2s+\alpha-2\right) +\lambda_{i, s}^{(+)} \right]^2 f_{s}^{( \alpha-2)}, \\
 f_{s}^{(1)}& = s+ \frac{1}{2} \left(\lambda_{i,s}^{(+)}-\lambda_{k}\right)+ \frac{\delta_{s 0}}{4}\lambda_{k}, \quad  f_{s}^{( 0)} = 1.
\end{align}
We do not have a closed-form solution to this recursion relation or a solution to the corresponding ODE for the generating function $\sum_{\alpha=0}^{\infty}x^{\alpha} f^{(\alpha)}_{s }$, although in practice the recursion relation itself is useful for finding the consistency conditions to a given order, as in the conformal bootstrap \cite{Poland:2018epd}. 

By equating these different channel decompositions for $\alpha =0, 1, \dots, \Lambda/2$, where $\Lambda$ is an even integer corresponding to the maximum number of derivatives, we obtain real consistency conditions that we can write in the following form:\footnote{To get a complete set when $s_1>0$, we start instead from $\alpha =1$ and add the consistency conditions from decomposing $\left\langle \nabla \phi^{(s_1)}_i \phi^{(s_2)}_j, \phi^{(s_1)}_i  \nabla \phi^{(s_2)}_j \right\rangle$. 
}
\begin{align} \label{eq:ij-crossed-ccs}
& V^{-1} \vec{F}''_0  +\sum_{k=1}^{\infty} \vec{F}'_{0}( \lambda_k) c^{(0, s_1, s_1)}_{k i i} c^{(0, s_2, s_2)}_{k j j}+\sum_{k=1}^{\infty}  \frac{1}{\chi_{J}(\lambda_k)} \vec{F}_{0}( \lambda_k) \left| c^{(0, s_1, s_2)}_{k i j} \right|^2 \nn \\
&+\sum^{s_2-s_1-\delta_{s_1 0}} _{n=1}\sum^{N_{g,n}}_{k=1}\vec{F}_{n} \left| c^{(n, s_1, s_2)}_{k i j} \right|^2 +\sum^{J} _{n=s_1+s_2}\sum^{N_{g,n}}_{k=1}\vec{F}_{n} \left| c^{(s_1, s_2, n)}_{i jk} \right|^2=0,
\end{align}
where $J\coloneqq \Lambda/2+s_1+s_2$ corresponds to the maximum exchanged spin (in physics terminology). The quantities $ \vec{F}'_{0}( \lambda_k)$ and $\vec{F}_{0}( \lambda_k)$ are vectors of polynomials of $\lambda_k$, while $\vec{F}''_0 $ and $\vec{F}_{n}$ are constant vectors---all of these vectors also depend in general on the fixed integers $s_a$ and the eigenvalue $\lambda_i$ if $s_1=0$.

A special case of the consistency conditions \eqref{eq:ij-crossed-ccs} is when there is a single fixed eigenmode, i.e.,  when $s_1=s_2$ and $i=j$ (we include here the case of a single  scalar eigenfunction). We can combine the mixed consistency conditions \eqref{eq:ij-crossed-ccs} with those for a single fixed eigenmode $\phi_i^{(s_1)}$ or $\phi_j^{(s_2)}$ to get a larger system of consistency conditions of the following form:
\begin{align}  \label{eq:ij-combined-ccs}
&
V^{-1} \vec{F}''_0 +
\sum_{k=1}^{\infty} \frac{1}{\chi_{J}(\lambda_k)}
\begin{pmatrix}
c_{k ii}^{(0, s_1, s_1)} \\
c_{k jj}^{(0, s_2, s_2)}
\end{pmatrix}  
 \vec{F}_0'( \lambda_k)
\begin{pmatrix}
c_{k ii}^{(0, s_1, s_1)} & c_{k jj}^{(0, s_2, s_2)}
\end{pmatrix} \nn \\
&+\sum_{k=1}^{\infty}\frac{1}{\chi_{J}(\lambda_k)} \vec{F}_{0}( \lambda_k) \left| c^{(0, s_1, s_2)}_{k i j} \right|^2 +\sum^{s_2-s_1-\delta_{s_1 0}} _{n=1}\sum^{N_{g,n}}_{k=1}\vec{F}_{n} \left| c^{(n, s_1, s_2)}_{k i j} \right|^2 
 +\sum^{J} _{n=s_1+s_2}\sum^{N_{g,n}}_{k=1}\vec{F}_{n} \left| c^{(s_1, s_2, n)}_{i jk} \right|^2 \nn \\
 & +\sum^{J}_{\substack{n=2 (s_1+\delta_{s_1 0}) \\ n \, \text{even}}} \sum^{N_{g,n}}_{k=1}\vec{F}'_{n} \left| c^{(s_1, s_1, n)}_{i ik} \right|^2+\sum^{J}_{\substack{n=2 s_2 \\ n \, \text{even}}} \sum^{N_{g,n}}_{k=1}\vec{F}''_{n} \left| c^{(s_2, s_2, n)}_{j jk} \right|^2 =0,
\end{align}
where $\vec{F}_0'( \lambda_k)$ is now a $2 \times 2$ matrix of polynomials of $\lambda_k$, $\vec{F}_{0}( \lambda_k)$ is a vector of polynomials of $\lambda_k$, and $\vec{F}_0''$, $\vec{F}_n$, $\vec{F}'_n$, and $\vec{F}''_n$ are constant vectors, with all of these vectors again depending on the fixed quantities $s_a$ and $\lambda_i$ (if $s_1$ vanishes). In Eq.~\eqref{eq:ij-combined-ccs}, we have included the consistency conditions of a single fixed eigenmode $\phi^{(s_a)}_a$ from quadruple overlap integrals with up to $2(J-2s_a)$ derivatives, which ensures that $J$ is the maximum exchanged spin for each contribution. This system of consistency conditions is easily generalised to include any number of fixed holomorphic $s$-differentials, and at most one scalar eigenfunction, by combining the single eigenmode consistency conditions for each fixed eigenmode with the mixed consistency conditions \eqref{eq:ij-crossed-ccs} for each pair of fixed eigenmodes; with $N$ fixed eigenmodes, $\vec{F}_0'( \lambda_k)$ is an $N \times N$ matrix.

\subsection{Bounds on the spectral gap}
\begin{subequations} \label{eqs:alpha-conditions}
To derive an upper bound on the spectral gap $\lambda_1$, we assume that $\lambda_1\geq \lambda^*$ for some fixed number $\lambda^*$ and then, for example, look for a linear combination of the consistency conditions \eqref{eq:ij-combined-ccs} with $s_1>0$ that leads to a contradiction. Specifically, if we have $n_{J}$ consistency conditions for some fixed positive integer $J$,  then we look for a vector $\vec{\alpha} \in \mathbb{R}^{n_J}$ such that the following conditions hold:
\begin{align}
& \vec{\alpha} \cdot  \vec{F}_0'(x) \succeq 0, \quad \forall x \geq \lambda^*, \\
&\vec{\alpha} \cdot  \vec{F}_0(x) \geq 0, \quad \forall x \geq \lambda^*,\\
&  \vec{\alpha} \cdot  \vec{F}''_0 = 1,  \\
& \vec{\alpha} \cdot  \vec{F}_n \geq 0, \quad n \in \{1, \dots, s_2-s_1 , s_1 +s_2, \dots, J\}, \\
& \vec{\alpha} \cdot  \vec{F}'_n \geq 0, \quad n \in \{2 s_1,2 s_1 +2, \dots , J\}, \\
& \vec{\alpha} \cdot  \vec{F}''_n \geq 0, \quad n \in \{2 s_2,2 s_2+2, \dots , J\},
\end{align}
\end{subequations}
where $\vec{\alpha} \cdot  \vec{F}_0'(x) \succeq 0$ means that $\vec{\alpha} \cdot  \vec{F}_0'(x)$ is a positive semidefinite matrix. If we can find such an $\vec{\alpha}$, then we reach a contradiction since $\vec{\alpha}$ dotted into the left-hand side of Eq.~\eqref{eq:ij-combined-ccs} gives a positive quantity, which cannot equal zero. We can then conclude that  our assumption that $\lambda_1\geq \lambda^*$ is inconsistent and hence $\lambda_1  < \lambda^*$ for any closed hyperbolic surface. We can repeat this procedure with different values of $\lambda^*$ to find the optimal bound for a given $J$. The optimal upper bound is nonincreasing with increasing $J$. A similar approach holds for a single fixed eigenmode or for more than two eigenmodes. The problem of finding a vector $\vec{\alpha}$ satisfying the conditions in Eqs.~\eqref{eqs:alpha-conditions} can be formulated as a semidefinite programming problem, following the approach of the conformal bootstrap \cite{Poland:2011ey, Kos:2014bka}. In this work, we use \texttt{SDPB} to numerically solve these semidefinite programs \cite{Simmons-Duffin:2015qma, Landry:2019qug}, using the nondefault settings listed in Table \ref{tab:SDPB}. 
\begin{table}[ht]
\centering
  \begin{tabular}{ |c | c| }
  \thickhline
\texttt{detectDualFeasibleJump} & True \\
\texttt{detectPrimalFeasibleJump} & True \\
\texttt{precision} & 1024 \\
\texttt{dualityGapThreshold} & $10^{-80}$ \\
\texttt{primalErrorThreshold} & $10^{-100}$ \\
\texttt{dualErrorThreshold} & $10^{-100}$ \\
\texttt{initialMatrixScalePrimal} & $10^{10}$ \\
\texttt{initialMatrixScaleDual} & $10^{10}$ \\
\texttt{maxComplementarity} & $10^{100}$\\
\thickhline
\end{tabular}
\caption{Our settings for \texttt{SDPB} version 2.5.1.}
\label{tab:SDPB}
\end{table}

Let us start with the case of a single fixed holomorphic $s$-differential. Taking $ J=30$, which corresponds to $30-2s+1$ consistency conditions, we find the upper bounds on the spectral gap shown in Table \ref{tab:single-cc-bounds} for $s \leq 6$. These bounds do not appear to be very competitive, since the Yang--Yau bound gives $\lambda_1 \leq 4$ for any closed hyperbolic surface. The reason is that these bounds also apply to more general spaces, such as quotients of surfaces by subgroups of their discrete isometry groups, since the consistency conditions are closed under the restriction to singlet eigenmodes. For the bounds in Table \ref{tab:single-cc-bounds} to apply to such a quotient, the original surface must admit a singlet holomorphic $s$-differential. For example, the smallest nondegenerate, nonzero eigenvalue of the Bolza surface is $\lambda_{24} \approx 23.07856$ \cite{Strohmaier_2012},\footnote{Although the first 23 nonzero eigenvalues appear to be degenerate eigenvalues to high numerical precision \cite{Strohmaier_2012}, in principle there could be accidental near degeneracies. See Refs.~\cite{Jenni1984, Cook-thesis} for work towards proving that the first two nontrivial eigenspaces have dimensions three and four, respectively.}
 which is consistent with Table~\ref{tab:single-cc-bounds} if the Bolza surface does not have a holomorphic differential or holomorphic quadratic differential transforming in the trivial representation of its isometry group. The proximity of $\lambda_{24}$ to the $s=3$ bound  in Table \ref{tab:single-cc-bounds} already suggests that the Bolza surface might be close to saturating certain  bootstrap bounds.
 \begin{table}[ht]
\centering
  \begin{tabular}{ c | c }
   $s$ & Upper bound on $\lambda_1$ \\ \thickhline
1 & 8.47032 \\
2 & 15.79144 \\
3 & 23.07916\\
4 & 30.35432  \\
5 & 37.62320\\
6 & 44.88836
\end{tabular}
\caption{Upper bounds on the spectral gap from consistency conditions of quadruple overlap integrals of a holomorphic $s$-differential with $J=30$, i.e., with $60-4s$ derivatives.}
\label{tab:single-cc-bounds}
\end{table}

To obtain a stronger upper bound, we should input additional information to forbid quotients with large gaps that are not closed surfaces. An approach that works well is to consider consistency conditions coming from quadruple overlap integrals with multiple fixed holomorphic $s$-differentials. Consider the mixed consistency conditions with two distinct holomorphic differentials, i.e., the consistency conditions in Eq.~\eqref{eq:ij-combined-ccs} with $s_1=s_2=1$ and $i \neq j$. This assumption is consistent with any closed hyperbolic surface since a genus-$g$ surface has $g$ independent holomorphic differentials and $g \geq 2$. For $J=50$, which corresponds to 196 consistency conditions, we obtain the numerical upper bound 
\be
\lambda_1 \leq 3.8388977.
\ee
This should be compared to the spectral gap of the Bolza surface $\lambda_1 \approx 3.8388873$ \cite{Strohmaier_2012}. Since the bootstrap bound is almost saturated by the Bolza surface, this provides strong evidence for the conjecture that the Bolza surface is the closed hyperbolic surface with the largest spectral gap \cite{Strohmaier_2012}. 

By considering a system of consistency conditions with $g_0$ distinct holomorphic differentials, we can obtain bounds that are valid for closed hyperbolic surfaces with genera $g \geq g_0$. We obtain the following results for $g_0 \leq 7$, which we compare to the best known bounds:
\begin{itemize}
\item For $g_0=3$, we find $\lambda_1 \leq 2.678483$ with $J=30$, corresponding to 261 consistency conditions.\footnote{We were aware of the analogous result of Ref.~\cite{Kravchuk:2021akc} when we first derived this bound.}
This is close to the spectral gap of the Klein quartic, which is the closed genus-$3$ surface with the largest symmetry group and has $\lambda_1 \approx 2.678$ \cite{Cook-thesis, Afkhami-Jeddi:2021qkf}.\footnote{The last digit here is deduced from results of the \texttt{FreeFem++} code of Ref.~\cite{Cook-thesis}, using meshes on the fundamental polygon with up to $n=110$ triangles along each side.}
The rigorous bound of Ref.~\cite{ros2021} gives  $\lambda_1 \leq 2(4- \sqrt{7}) \approx 2.71$ for $g=3$. 
\item For $g_0 = 4$, we find $\lambda_1 \leq 2.17$ with $J=10$, corresponding to 144 consistency conditions. This is weaker than the Yang--Yau bound, which gives $\lambda_1\leq 2$ for $g= 4$. The most symmetric closed hyperbolic surface of genus 4 is Bring's surface, which has $\lambda_1 \approx 1.9$ \cite{Cook-thesis}.  
\item For $g_0=5$, we find $\lambda_1 \leq 1.86$ with $J=10$, corresponding to 225 consistency conditions. The rigorous bound of Ref.~\cite{karpukhin2021} gives $\lambda_1 \leq(47-\sqrt{977})/8 \approx 1.97$ for $g=5$.
\item For $g_0=6$, we find $\lambda_1 \leq 1.67$ with $J=10$, corresponding to 324 consistency conditions. This is weaker than the Yang--Yau bound, which gives $\lambda_1\leq 1.6$ for $g= 6$. 
\item For $g_0=7$, we find $\lambda_1 \leq 1.55$ with $J=10$, corresponding to 441 consistency conditions. The rigorous bound of Ref.~\cite{karpukhin2021} gives $\lambda_1 \leq(33-\sqrt{543})/6 \approx 1.62$ for $g=7$. 
\end{itemize}
These results show that bootstrap methods can also produce strong bounds for surfaces of larger genera.

\section{Conclusions}
\label{sec:Conclusions}

We have shown how to derive consistency conditions for closed hyperbolic surfaces coming from quadruple overlap integrals with arbitrarily many derivatives and involving holomorphic $s$-differentials. These improve on the explicit consistency conditions for closed hyperbolic manifolds found in Ref.~\cite{Bonifacio:2021msa}, which had fixed scalar eigenfunctions and at most 16 derivatives, albeit at the cost of restricting to two dimensions. With these additional consistency conditions, it is possible to derive more powerful bootstrap bounds. One example we showed was a numerical upper bound on the spectral gap of a closed orientable hyperbolic surface that is stronger than the Yang--Yau bound for genus 2 and is nearly saturated by the Bolza surface.

There is much more that could be investigated in the future. In two dimensions, one could try to learn more about the closed hyperbolic surfaces that are close to saturating bootstrap bounds, e.g., by using the extremal functional method from the conformal bootstrap \cite{El-Showk:2012vjm}. It would be interesting to try to develop analytic bootstrap techniques and to explore the large-genus regime. 
It would also be interesting to consider consistency conditions coming from quadruple overlap integrals of tensors in higher dimensions, especially in three dimensions.

\paragraph{Acknowledgements:} I would like to thank Anthony Ashmore, Kurt Hinterbichler, and Alex Maloney for helpful discussions. I am supported by the research program VIDI with Project No. 680-47-535, which is partly financed by the Netherlands Organisation for Scientific Research (NWO). This work has been partially supported by STFC consolidated grant ST/T000694/1.

\renewcommand{\em}{}
\bibliographystyle{utphys}
\addcontentsline{toc}{section}{References}
\bibliography{hyperbolic-refs}

\end{document}